\documentclass[aps,pra,showpacs, superscriptaddress,twocolumn]{revtex4-1}

\usepackage{graphicx,amsmath,amsfonts,hyperref}
\usepackage{color}
\usepackage{soul}
\usepackage{braket}
\usepackage{bm} 

\begin{document}


\title{Quantum to classical transition induced by gravitational time dilation}

\author{Boris Sokolov} 
\affiliation{Turku Center for Quantum Physics, Department of Physics and
Astronomy, University of Turku, FIN-20014 Turku, Finland}
\email[]{bosoko@utu.fi} \homepage[]{www.tqt.fi}

\author{Iiro Vilja}\affiliation{Turku Center for Quantum Physics, Department of Physics and Astronomy, University of Turku, FIN-20014 Turku, Finland}

\author{Sabrina Maniscalco}\affiliation{Turku Center for Quantum Physics, Department of Physics and Astronomy, University of Turku, FIN-20014 Turku, Finland}\affiliation{Centre for Quantum Engineering, Department of Applied Physics, School of Science, Aalto University, P.O. Box 11000, FIN-00076 Aalto, Finland}

\date{\today}

\begin{abstract}
We study the loss of quantumness caused by time dilation \cite{Brukner} for a Schr\"odinger cat state. We give a holistic view of the quantum to classical transition by comparing the dynamics of several nonclassicality indicators, such as the Wigner function interference fringe, the negativity of the Wigner function, the nonclassical depth, the Vogel criterion and the Klyshko criterion. Our results show that only two of these indicators depend critically on the size of the cat, namely on how macroscopic the superposition is. Finally we compare the gravitation-induced decoherence times to the typical decoherence times due to classical noise originating from the unavoidable statistical fluctuations in the characteristic parameters of the system \cite{CollapseRevival}. We show that the experimental observation of decoherence due to time dilation imposes severe limitations on the allowed levels of classical noise in the experiments. 
\end{abstract}


\maketitle
\section{Introduction}
In presence of massive objects, the geometry of spacetime is warped. This warping, commonly called the curvature of spacetime, causes clocks situated at different locations to tick at different rates. This is known as gravitational time dilation and has been an experimentally proven phenomenon for about half of a century \cite{PoundRebka}. Even the relatively weak gravitational field of the Earth has an effect on certain technological applications such as the Global Positioning System and must be accounted for.

Modern experiments \cite{ScienceOpticalClocks} have been able to detect time dilation near the Earths surface due to a height difference of 1 meter. Furthermore, time dilation was theoretically predicted to cause phase shift as well as loss of visibility measurable as a change in the interference pattern in appropriately designed interferometers \cite{SinhaSamuel, MullerPetersChu2010, Zych2011}, potentially increasing the measurement precision.

Recently, Pikovski et al. have considered the effect of gravitational time dilation on the coherence of a composite quantum system using the tools of open quantum systems theory. A recent review on the subject can be found in Ref. \cite{PikovskiReview2017}. The quantum system is prepared in a quantum superposition of two locations corresponding to different distances from Earth \cite{Brukner}. For this system, e.g., a molecule, time dilation induces a universal coupling between the internal degrees of freedom and the center of mass (c.m.). This leads to decoherence in the c.m. position of the particle characterized by a time scale $\tau_{\rm dec}$. From an open quantum systems perspective, the internal degrees of freedom of the molecule play the role of the environment while the c.m. degree of freedom represents an open system. Note, that the total gravitational c.m. weight consists of the total mass and the internal energy as both act as sources of the gravity in general relativity; this is incorporated in Ref. \cite{Brukner} as well as in the present study. 

For the sake of concreteness, let us consider the case in which the internal degrees of freedom are bosonic. A standard example is a molecule for which the vibrations of its constituent atoms are described in terms of $N$ normal modes represented mathematically as uncoupled quantum harmonic oscillators with frequencies $\omega_i$, $i \in \left\{0,1,..,N\right\}$. The coupling between system and environment arises because the vibrational frequencies become position-dependent due to time dilation, namely, $\omega_i \rightarrow \omega_i(x)$.
This in turn induces non-dissipative decoherence, as shown in Ref. \cite{Brukner,bassi1,bassi2}. 

The gravitation-induced decoherence model considered by Pikovski et al. relies on a number of quite restrictive approximations and should therefore be considered as a toy model useful to grasp the main features of a new decoherence mechanism. More specifically, the spherically symmetric metric is considered in the Newtonian limit, valid only for slowly moving particles in weak gravitational fields, therefore general relativistic effects are not taken into account. Secondly, a superposition of position eigenstates is considered as initial state, rather than a more physically significant superposition of macroscopically distinguishable classical-like states. Thirdly, the observed system is assumed to be in the same relativistic frame as the observer. These assumptions may severely limit the generality of the conclusions made in Ref. \cite{Brukner}, and it is therefore crucial to extend the proposed model towards more realistic physical scenarios.

In this paper we generalize the study of gravitationally induced decoherence in several directions. First we focus on the case in which the molecule is confined in a harmonic trap and initially prepared in a superposition of two macroscopically distinguishable coherent states of the c.m. motion, centered at two different heights. This allows us to describe gravitation-induced decoherence when the model open system is a quantum harmonic oscillator. This is particularly useful in describing the quantum to classical transition and, in fact, experiments revealing the decoherence of these so called cat states in this framework have been performed in different experimental platforms \cite{Haroche,Wineland}. 

Importantly, the nonclassicality of this Schr\"odinger cat state can be described by means of various, physically meaningful nonclassicality indicators, i.e., the peak of the interference fringe of the Wigner function \cite{InterFringePeak}, the nonclassical depth \cite{nonclassdepth,nonclassdepth2}, the negativity of the Wigner function \cite{WignerNeg}, the Vogel criterion \cite{Vogel} and the Klyshko \cite{Klyshko_paper} criterion. Such quantities measure different, nonequivalent ways in which our system can be nonclassical. Therefore, a much more complete picture of the quantum to classical transition can be given by looking at the dynamics of all such indicators, since in general their characteristic decoherence timescales will be different. This is one of the main goals of this paper.

Finally, we take into account realistic sources of noise arising in the experiments. We compare the characteristic timescales of time dilation induced decoherence to the timescale of decoherence due to classical fluctuation in crucial experimental parameters. This allows us to assess whether time dilation induced decoherence is an observable phenomenon or, alternatively, to define the necessary measurement precision for the detection of such an intriguing fundamental effect.
%

The paper is structured as follows. In Sec. \ref{sec:model} we present the master equation and its exact solution in terms of the quantum characteristic function. In Sec. \ref{sec:decoh_dynamics}, we study the dynamics of an initial superposition of coherent states and describe decoherence in terms of the time evolution of the interference fringe of the Wigner function. In Sec. \ref{sec:nonclass_ind}, we calculate, estimate and compare the values of the aforementioned nonclassicality indicators for our system. Finally, in Sec. \ref{sec:classic_noise}, we consider the effect of classical noise and estimate the levels of precision and noise control required to observe the time dilation induced decoherence phenomenon.

\section{The master equation and its solution}\label{sec:model}

Following Ref. \cite{Brukner} we consider a composite system described by the subsequent Hamiltonian

\begin{equation}\label{eq:fullHamil}
H = H_{\mathrm{cm}} + H_0 + H_{\text{int}},
\end{equation}

where
\begin{align}
H_0 &= \sum_{i=1}^N \hbar \omega_i n_i, \label{eq:fullHamilH0}\\
H_{\mathrm{int}} &= \hbar \frac{g x}{c^2} \left( \sum_{i=1}^N \omega_i n_i\right), \label{eq:fullHamilHint} 
\end{align}
with $H_0$ the Hamiltonian of the internal degrees of freedom described by $N$ quantum harmonic oscillators of frequencies $\omega_i$. 
For weak gravitational fields and slow-moving particles \cite{Note1}, the time dilation induced coupling between the internal degrees of freedom and the c.m. to the lowest order in $c^{-2}$ is described by $H_{\mathrm{int}}$. In Eq. (\ref{eq:fullHamilHint}), $g$ is the local gravitational acceleration and $x$ is the c.m. position operator. Using standard open quantum systems approaches \cite{BreuerPetruccione} one derives the master equation for the c.m. dynamics under the following assumptions: (i) weak coupling limit, (ii) no initial correlations between the state of the c.m. and the state of internal degrees of freedom, (iii) initially thermalized state of the internal degrees of freedom, (iv) negligible changes in the off-diagonal elements due to the c.m. Hamiltonian on the decoherence timescale \cite{Brukner}:

\begin{equation}\label{eq:meBruk}
\begin{split}
\dot{\rho}_{\text{cm}}(t) = &- \frac{i}{\hbar} \left[ \tilde{H}_{\text{cm}} + \left(  m+ \frac{\bar{E}_0}{c^2}\right) gx, \rho_{\text{cm}}(t)\right] \\
&- \left( \frac{\Delta E_0 g}{\hbar c^2} \right)^2 t \left[ x, \left[ x,\rho_{\text{cm}}(t)\right] \right],
\end{split}
\end{equation}

In Eq. (\ref{eq:meBruk}), $m$ is the total mass of the system while $\bar{E}_0$  and $\Delta E_0^2 = \langle H_0^2 \rangle - \langle H_0 \rangle ^2$  are the expectation value and the variance of $H_0$, respectively. The term $\tilde{H}_{\text{cm}}$ represents the c.m. Hamiltonian in a convenient picture, as defined in Ref. \cite{Brukner}. 

We now consider the case where the c.m. motion is a quantum harmonic oscillator $H_{\text{cm}} = \frac{\hbar \omega_0}{2} a^\dag a$, with $a^\dag$ and $a$ creation and annihilation operators, respectively, and $\omega_0$ the oscillator frequency. 
Assuming that the dominant term in the unitary dynamics is the c.m. Hamiltonian,  Eq. (\ref{eq:meBruk}) becomes

\begin{equation}\label{eq:meBrukShort}
\dot{\rho}_{\text{cm}}(t) =- \frac{i}{\hbar} \left[ H_{\text{cm}} , \rho_{\text{cm}}(t)\right] -\Delta(t) \left[ X, \left[X,\rho_{\text{cm}}(t) \right] \right],
\end{equation}
where we have used the dimensionless position operator $X =  \frac{1}{\sqrt{2}}(a^\dag + a)$,  related to $x$ through the formula $x=\sqrt{\frac{\hbar}{ m \omega_0}} X \equiv \Delta x_0 X$, with $\Delta x_0$  the width of the ground state wavefunction of the quantum harmonic oscillator.
With this notation the time-dependent coefficient $\Delta(t)$, which is linear in time and positive, takes the form
\begin{equation}\label{eq:delta}
\Delta(t) =  \left( \frac{\hbar}{m \omega_0} \right) \left( \frac{\Delta E_0 g}{\hbar c^2}\right)^2 t \equiv \kappa t.
\end{equation}


Note that the master equation  (\ref{eq:meBrukShort}) is of the Lindblad form hence the dynamics is always divisible (Markovian) \cite{DarekSab}. 
Moreover, the master equation here considered is a special case of the well-known quantum Brownian motion master equation which, in the interaction picture, reads as follows \cite{BreuerPetruccione}
\begin{equation}
\begin{split}
\frac{d \rho(t)}{dt} =& - \Delta(t) \left[ X, \left[ X, \rho(t) \right]\right]  \\
&+ \Pi(t) \left[ X,\left[ P,\rho(t)\right]\right] + \frac{i}{2} r(t) \left[ X^2 , \rho(t)\right]  \\
&- i \gamma(t) \left[ X, \left\{P, \rho(t) \right\}\right]. 
\end{split}
\end{equation}
Indeed, the master equation above reduces to  Eq. (\ref{eq:meBrukShort}) for $\Pi(t), r(t), \gamma(t) = 0$.
Using the general solution of the quantum Brownian motion model (see, e.g., Refs. \cite{DecControl, FiniteTimeTransitionCat, NonMarkLimitBrownian, Misbeliefs}) we obtain the following simple solution for $\rho_{\text{cm}}(t)$, in terms of the symmetrically ordered quantum characteristic function $\chi_t(\xi)$
\begin{equation}\label{eq:phi_with_ctdt}
\chi_t(\xi) = \chi_0( \xi) e^{-N(t) |\xi|^2},
\end{equation}
with $\xi \in \mathbb{C}$ and
\begin{equation} \label{nofteq}
N(t) = \frac{1}{2} \kappa t^2 = \frac{1}{2} \left( \frac{\Delta E_0 g}{\hbar c^2}\right)^2 \left( \frac{\hbar}{m \omega_0} \right) t^2.
\end{equation}
%
We introduce here for convenience the family of $s$-ordered characteristic functions defined as
\begin{equation} 
\chi(\xi,s) = \mathrm{Tr}\left[ \rho  D(\xi) \right] e^{\frac{1}{2} s |\xi|^2},
\end{equation}
where $D(\xi)=e^{\xi \hat{a}^\dag - \xi^* \hat{a}}$ is the displacement operator and $s=1,0,-1$, correspond to the normally, symmetrically and antinormally ordered characteristic functions, respectively.  The quasi-probability distributions $W(\alpha,s)$, defined as the Fourier transform of the quantum characteristic functions,
\begin{eqnarray} \label{eq:charfunctionxi}
W(\alpha,s) &= \frac{1}{\pi^2} \int d^2 \xi  e^{\alpha \xi^* - \alpha^* \xi} \chi(\xi,s),
\end{eqnarray}
are the well-known Wigner ($s=0$), Glauber or $P$ ($s=1$) and Husimi or $Q$ ($s=-1$) functions, respectively, extensively used in quantum optics. In the following we use, for simplicity, the following notation $\chi_t(\xi,0) \equiv \chi_t(\xi)$ and $W(\alpha,0) \equiv W(\alpha) $. 

The description in terms of characteristic function is completely equivalent to the density operator formalism as one sees from the equation
\begin{eqnarray} \label{rhoxi}
\rho(t)=\frac{1}{2 \pi} \int d\xi d\xi^* \chi_t(\xi) D(\xi).
\end{eqnarray}
The solution of the master equation (\ref{eq:meBruk}) is therefore given by Eq. (\ref{rhoxi}) with Eqs. (\ref{eq:phi_with_ctdt}) -(\ref{nofteq}).


\section{Decoherence dynamics}\label{sec:decoh_dynamics}

Let us assume that the system is initially prepared in a Schr\"odinger cat state of the form 
\begin{equation}
\rho_{\text{cm}} (0) = | \Psi \rangle  \langle \Psi |,
\end{equation}
where $| \Psi \rangle = 
\frac{1}{\sqrt{\mathcal{N}}}\left( |\alpha \rangle + | - \alpha \rangle\right)$, with $| \alpha \rangle$  a coherent state. Here, $\mathcal{N}^{-1}=2 \left( 1 + e^{-2|\alpha|^2}\right)$ is the normalization factor. More specifically we consider the so-called even coherent state obtained for $ \alpha \in \mathbb{R} $.
The "size" of the cat is given by $\Delta x = 2 \alpha \Delta x_0$, where $2 \alpha$ is the distance between the peaks of the two Gaussian functions describing the coherent state components of the superposition in phase space.

%
%
%
%

We study the time evolution of the state in terms of its Wigner function defined in Eq. (\ref{eq:charfunctionxi}) with $s=0$. Inserting Eqs.(\ref{eq:phi_with_ctdt})-(\ref{nofteq}) into (\ref{eq:charfunctionxi}) one can write the Wigner function for the initial state here considered as follows  \cite{NonMarkLimitBrownian}:


\begin{equation}
W(\beta,t)= W^{+\alpha}(\beta,t)+W^{-\alpha}(\beta,t)+W_I(\beta,t),
\end{equation}
where
\begin{equation}
\begin{split}
W^{\pm \alpha}(\beta,t) =& \frac{\mathcal{N}}{\pi (N(t) + 1/4)^{1/2}}\exp\left( -\frac{\mathrm{Im}(\beta)^2}{2 N(t) + 1/2} \right)\\
&\times \exp\left( - \frac{(\mathrm{Re}\beta \mp \alpha)^2}{1/2}\right)\text{, and}
\end{split}
\end{equation}
\begin{equation}
\begin{split}
W_I(\beta,t) =& \frac{2\mathcal{N}}{\pi (N(t) + 1/4)^{1/2}} \cos \left( \frac{2}{2N(t) + 1/2}\alpha \mathrm{Im}\beta \right) \\
&\times \exp\left( - 2 \alpha^2 \left( 1- \frac{1}{4N(t) + 1} \right) \right) \\
&\times \exp\left( -\frac{\mathrm{Im}(\beta)^2}{2N(t) + 1/2} -\frac{\mathrm{Re}(\beta)^2}{1/2} \right)
\end{split}
\end{equation}

with  $N(t)=\int_0^t dt' \Delta(t')$ given by Eq. (\ref{nofteq}).

The expression above is particularly suited to describe environment-induced decoherence and its effect on the cat state since it singles out the interference term $W_I(\beta,t)$ which characterizes the quantumness of the initial state. As decoherence takes place such term decays with a characteristic time known as decoherence time. Following the standard description of environment-induced decoherence, we introduce the fringe visibility function \cite{PazHabibZurek}
\begin{align}
F(\alpha,t) = \frac{1}{2} \frac{W_I(\beta,t)|_{peak}}{\left[W^{+\alpha}(\beta,t)|_{peak} W^{-\alpha}(\beta,t)|_{peak}\right]^{1/2}},
\end{align}
where 
\begin{equation}
\begin{split}
W_I(\beta,t)|_{peak} &= W_I(\beta=(0,0),t) \\
W^{\pm \alpha}(\beta,t)|_{peak} &= W^{\pm \alpha }(\beta=(\pm \alpha,0),t).
\end{split}
\end{equation}

%

In our case, this function takes the form \cite{DecControl, FiniteTimeTransitionCat, NonMarkLimitBrownian, Misbeliefs} 
\begin{equation}\label{eq:decorateManiscalco}
F(\alpha,t) = \exp \left[ -2 \alpha^2 \left( 1-\frac{1}{1+4N(t)}\right) \right].
\end{equation}

%
%
%
%
Inserting Eq. (\ref{nofteq}) into Eq. (\ref{eq:decorateManiscalco}), and in the limit $\kappa t^2 \ll 1$, we obtain the following simple expression for the fringe visibility function 
\begin{equation} \label{fapproxim}
F(\alpha,t) \approx e^{-2 \alpha^2 \kappa t^2} \equiv e^{-(t/\tau_{\text{dec}})^2},
\end{equation}
where the decoherence time is given by
\begin{equation}
\label{eq:decorate_fringepeak}
\tau_{\text{dec}}^2 = \frac{1}{{4 \alpha^2 \kappa}} 
\end{equation}
The condition $\kappa t^2 \ll 1$ can be written as $t \ll \tau_{\text{dec}} \Delta x/ \Delta x_0$ showing that Eq. (\ref{fapproxim}) only describes the initial decoherence behaviour since generally $\Delta x/ \Delta x_0 >1$.


To conclude this section we compare the decoherence time derived with our formalism with the one defined in Ref. \cite{Brukner}, where the initial state of the c.m. is a superposition of different position eigenstates separated by a height difference $\Delta x$ and decoherence is measured using the notion of interferometric visibility given by \cite{Brukner}:
\begin{equation}
V(t) \approx \left(1+\left(k_B T g \Delta x \frac{t}{\hbar c^2}\right)^2\right)^{-N/2},
\end{equation}
with $N$ the number of internal degrees of freedom. For times $t^2 \ll N \bar{\tau}_{\text{dec}}^2$, with the decoherence time defined as
\begin{equation}
\bar{\tau}_{\text{dec}}=\sqrt{\frac{2}{N}} \frac{\hbar c^2}{k_B T g \Delta x},
\end{equation}
the interferometric visibility is approximated by
\begin{equation}\label{eq:decorateBrukner}
V(t) \approx e^{-\left( t/\bar{\tau}_{\text{dec}}\right)^2},
\end{equation}
which has the same temporal behaviour as the Wigner function fringe visibility that we have derived in Eq. (\ref{fapproxim}). Recalling from Eq. (\ref{nofteq}) the expression of $\kappa$ appearing in Eq. (\ref{eq:decorate_fringepeak}), and remembering that $\Delta E_0^2 = N k_B^2 T^2$ when the state of the internal degrees of freedom is thermal,   one concludes straightforwardly that $\bar{\tau}_{\text{dec}} = \tau_{\text{dec}}$.

Note, that in general both the system and the observer are at finite distances from the source of gravity, say $r_{\text{sys}}$ and $r_{\text{obs}}$. In most experimental settings one can assume that these two distances practically coincide, since the observer performing the measurement is in the laboratory where the system is. However, one can imagine, e.g., a setting for which the laboratory is on a satellite and the data are sent to a measuring observer on Earth. This implies that the standard relative dilation factor relates the coherence times of the corresponding rest frames
\begin{equation}
- \left( 1 - \frac{r_s}{r_{{\text{sys}}}} \right)^{-1} dt_{\text{sys}}^2 = - \left(1 - \frac{r_s}{r_{\text{obs}}}\right)^{-1} dt_{\text{obs}}^2,
\end{equation}
with $r_s$ the Schwarzschild radius. Thus the decoherence time measured in observer's rest frame is
\begin{equation}
\tau_{\text{\text{dec}}}^{\text{obs}} \equiv \frac{a_{\text{obs}}}{a_{\text{sys}}}\,\tau^{\text{sys}}_{\text{\text{dec}}} 
\end{equation}
where $\tau^{\text{sys}}_{\text{\text{dec}}}$ is the decoherence time in the rest frame of the system and $a_{{\text{sys}},{\text{obs}}} = \sqrt{1- \frac{r_{s}}{r_{{\text{sys}},{\text{obs}}}}} $.
This carries the lowest order gravitational effects to the complete quantum system-observer pair. For Earth mass and radius the correction $(\tau_{\text{\text{dec}}}^{\text{obs}}-\tau_{\text{\text{dec}}}^{{\text{sys}}})/\tau_{\text{\text{dec}}}^{\text{obs}} $ is at most only of the order of $10^{-13}$ at surface level system. Therefore we drop the superscript $\text{obs}$ systematically from the decoherence time. However, this relativistic affect is more significant in proximity to heavy stellar object or near a horizon.

\section{Dynamics of nonclassicality indicators}\label{sec:nonclass_ind}

The definition of nonclassicality for the states of the quantum harmonic oscillator has been extensively debated in the past. There exist indeed different quantities measuring or highlighting different ways in which this paradigmatic system departs from classical behaviour. Therefore, in order to provide a holistic view of the quantum to classical transition stemming from gravitational decoherence, in this section we explore how the most widespread nonclassicality indicators witness the loss of quantumness.

We will describe the dynamics of the following quantities: the nonclassical depth, measuring the minimum number of thermal photons required to destroy any nonclassical characteristics of the system \cite{nonclassdepth}; the negativity of the Wigner function quantifying the separation between the Wigner distribution and a classical probability distribution \cite{WignerNeg}; the Vogel criterion defined in terms of properties of the characteristic functions of the quadrature distributions having no classical counterpart \cite{Vogel}; and the Klyshko criterion detecting differences between classical photon number probability distributions and the quantum ones \cite{Klyshko_paper}.



Let us begin by writing the quasiprobability distributions defined in Eq. (\ref{eq:charfunctionxi}) for $s=-1,0,1$ as the following convolution
\begin{equation}
\begin{split} \label{eq:genw}
W(\alpha,s) &= W(\alpha,s') \star G(s'-s,\alpha) \\
&= \int d^2 \beta W(\beta,s') G(s'-s,\alpha-\beta),
\end{split}
\end{equation}
where
\begin{equation}
G(s'-s,\alpha) = \frac{2}{\pi \left( s'-s \right)} e^{-2\frac{|\alpha|^2}{s'-s}}.
\end{equation}
We now generalize to the case in which $s$ is a continuous parameter taking values in the interval $s\in [-1,1]$. 

The \emph{nonclassical depth} is defined as \cite{nonclassdepth}
\begin{equation}
\begin{split}
\eta &= \frac{1}{2} (1-\bar{s}), \\
\bar{s}&= \sup \left\{ s \in [-1,1] \left| W(\alpha,s) \geq 0 \right. \right\}.
\end{split}
\end{equation}

By following the steps of Ref. \cite{FiniteTimeTransitionCat} we notice that setting $s'=1$ in Eq. (\ref{eq:genw}) we can express $W(\alpha,s)$ as a convolution of the $P$ function
\begin{equation}
W(\alpha,s) = P (\alpha) \star G(1-s,\alpha).
\end{equation}
Using Eqs.  (\ref{eq:phi_with_ctdt})-(\ref{nofteq}) we obtain the following expression for the time evolution of the $P$ function
\begin{eqnarray}
P_t (\alpha) &=& \frac{1}{\pi} \int d \xi^2 \chi_0(\xi) e^{-N(t) |\xi|^2+\alpha \xi^* -\alpha^* \xi}  \nonumber \\
&=& P_0(\alpha) \star G(1-s_t,\alpha)  = W(\alpha,s_t), \label{eqkey}
\end{eqnarray}
where we have used the fact that the Fourier transform of a product of two functions is equal to the convolution of the two corresponding Fourier transforms, and where $s_{t} = 1 - 2 N(t)$. The equation above shows that our dynamics transform the initial $P$ function into the other characteristic functions. Since the nonclassical depth of the initial state is $\eta = 1$, one can prove that the time $\tau_p$ at which $W(\alpha,s)$ becomes positive corresponds to the time at which the initial $P$ is transformed into the $Q$ function (which is always positive) \cite{CollapseRevival},  that is 
\begin{equation}\label{eq:st-1_equation}
\begin{split}
s_{\tau_p} &\equiv 1 - 2 N(\tau_p) = -1.
\end{split}
\end{equation}
Solving for $\tau_p$ one obtains straightforwardly
\begin{equation}
\tau_p^2 = 2/\kappa.
\end{equation}
Note that $\tau_p = 2 \alpha \tau_{\text{dec}}$, where $\tau_{\text{dec}}$ is the decoherence time associated to the decay of the Wigner function fringe visibility, given by Eq. (\ref{eq:decorate_fringepeak}). Hence, $\tau_p$ does not depend on the size of the cat, contrarily to $\tau_{\text{dec}}$. Moreover, for truly macroscopic superpositions such that $|\alpha|>>1$, the loss of interference in the Wigner function is much faster than the loss of quantumness measured by the nonclassical depth.

We now turn our attention to the second nonclassicality indicator, namely the  \emph{negativity of the Wigner function}. More precisely, we are interested in identifying the time $\tau_W$ at which the Wigner function of the initial cat state, which is negative in several zones of the phase space, becomes positive everywhere. This time can be calculated analytically once again using Eq. (\ref{eqkey}) and corresponds to the time at which the initial $P$ function is turned into the Wigner function
\begin{equation}
s_{\tau_W} = 1- 2 \kappa \tau_W^2 = 0,
\end{equation}
yielding
\begin{equation}\label{eq:tau_w}
\tau_W^2=1/\kappa = \tau_p^2/2.
\end{equation}
Decoherence induced by time dilation therefore causes the negativity of the Wigner function to disappear faster than the nonclassicality as measured by the nonclassical depth. Moreover, as the latter one, it does not depend on the size of the initial superposition.

\begin{figure}
\vspace{1cm}
\includegraphics[width=0.48\textwidth]{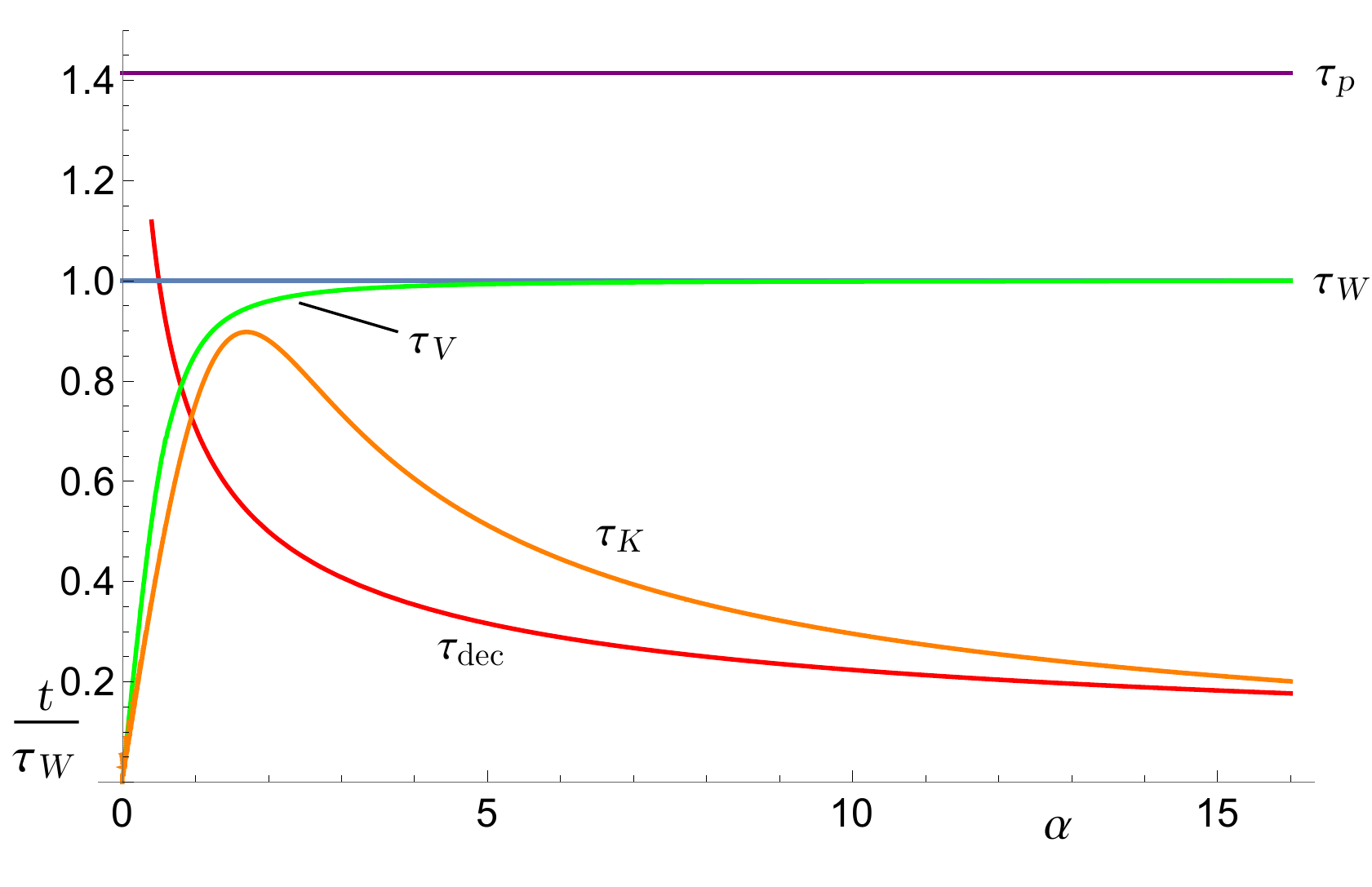}
\caption{(colors online) Behavior of different nonclassicality indicators plotted as a function of (dimensionless) $\alpha$ in units of $\tau_W$ (seconds). $\tau_{\text{\text{dec}}}$ represents the peak of interference fringe.}\label{fig:criterionplot}
\end{figure}

The two criteria considered so far are based on properties of the quasiprobability distribution functions and, as such, are experimentally demanding since they require full tomography of the state while it evolves due to the interaction with the environment. The next criterion examined is on the contrary experimentally easier to implement since it is defined in terms of the symmetrically ordered characteristic function which can be directly measured through balanced homodyne detection. The \emph{Vogel nonclassicality criterion} is indeed simply defined as follows \cite{Vogel}: a state is nonclassical at time $t$ iff for its normally ordered characteristic function,
\begin{equation}
\exists u,v \in \mathbb{R} \text{ s.t. } |\chi_t(\xi, 1)| > 1 \text{, with } \xi = u+iv.
\end{equation}
Using our solution given by Eq. (\ref{eq:phi_with_ctdt}) one promptly obtains
\begin{equation}\label{eq:phi_with_ctdt3}
\chi_t(\xi,1) = \chi_0(\xi,1) e^{-N(t) |\xi|^2}.
\end{equation}
Recalling the expression of $\chi_0(\xi,1)$ for our initial Schr\"odinger cat state we can write the Vogel criterion as follows
\begin{equation}\label{eq:phicharfunc}
\begin{split}
\chi_{t}(u,v,1)&>1  , \text{ where}\\
\chi_t(u,v,1) &= \frac{2}{\mathcal{N}}e^{-N(t)(u^2+v^2)}\left[ \cos(2 \alpha v) \right.\\ 
&+ \left.  e^{-2\alpha} \cosh(2 \alpha u) \right] \le \chi_t(u,0,1).
\end{split}
\end{equation}
The Vogel criterion is inherently state-dependent and, in particular, in our case it depends on $\alpha$. One defines the Vogel nonclassicality transition time $\tau_V$ from the equation $\chi_{\tau_V}(u,0,1)=1$. 
In Fig. \ref{fig:criterionplot} we plot the behaviour of $\tau_V$, in units of $\tau_W$, as $\alpha$ increases. As one can see from the figure, the decoherence time $\tau_V$ tends to saturate as the superposition becomes more and more macroscopic. In the limit $\alpha \rightarrow \infty$ one easily obtains analytically that the condition $\chi_{\tau_V}(u,0,1)=1$ amounts at requiring that 
\begin{equation}
 2 N(\tau_V) \ge 1 \Rightarrow \tau_V^2 = 1/\kappa = \tau_W^2.
\end{equation}
This is an upper bound for the onset of classicality and, for our system, it turns out to be equivalent to the disappearence of the negativity of the Wigner function. 

The last criterion explored in our paper is the \emph{Klyshko criterion} \cite{Klyshko_paper}. Similarly to the Vogel criterion, also this one is experimentally friendly since it only requires the measurement of phonon number distributions $p(n)=\langle n|\rho|n\rangle$. The criterion states that a sufficient condition for a state to be nonclassical, or more precisely to have nonclassical phonon counting statistics, is that there exist at least one integer $n$ such that \cite{DAriano,Klyshko_paper,FiniteTimeTransitionCat}
\begin{equation}\label{eq:b(n)}
B(n) \equiv (n+2) p(n) p(n+2) - (n+1) (p(n+1))^2 < 0.
\end{equation}

The phonon number probabilities can be written in terms of the normally ordered characteristic functions $\chi_t(u,v,1)$ and the antinormally ordered characteristic function for the number states $\chi_n(u,v,-1)=e^{-u^2-v^2}L_n(u^2+v^2)$, where $L_n(x)$ are the Laguerre polynomials:
\begin{equation}
p(n,t)= \frac{1}{\pi} \int du dv \chi_t (u,v,1) e^{-u^2-v^2}L_n(u^2+v^2).
\end{equation}

Finding times where $B(1)=0$ is valid yields the plot in Fig. \ref{fig:criterionplot}. The plot shows that the Klyshko nonclassicality decoherence time $\tau_K$ (in the figure in units of $\tau_W$) depends on the size of the cat and, in particular, quickly decreases for increasing values of $\alpha$, i.e., with the size of the cat state.

We also see that, in general, $\tau_K < \tau_V < \tau_W < \tau_p$ so, as in the case of $\tau_{\text{dec}}$, this type of nonclassicality quickly disappears and it is more difficult to be hidden by other sources of noise.

In order to evaluate the ability to detect time dilation induced decoherence we need to compare its characteristic time scales with those of the most significant noise sources that may affect the dynamics. The fact that gravitational decoherence might be very hard to detect experimentally was already recognized in Ref. \cite{Brukner}, where the effect of decoherence due to emission of radiation was considered for comparison. A more thorough analysis was performed Ref. \cite{bassi2}, where the effects of both collisional and thermal decoherence were analyzed. In the following section we will perform a comparison with what is perhaps the most common source of decoherence in the experiments, namely classical noise on the experimental parameters.

\section{Comparison between time dilation induced decoherence and classical noise}\label{sec:classic_noise}

One of the most ubiquitous sources of decoherence in interferometric experiments is classical noise affecting the relevant experimental parameters. The generality of this noise source makes the investigation of its effects on the new type of decoherence described in this paper a crucial step. Moreover, contrarily to the environmental effects considered before in Refs. \cite{Brukner} and  \cite{bassi2}, in the case of classical stochastic noise it is possible to perform a numerical exact analysis without invoking the Born/Markov approximations commonly done for quantum environments. In this way we can extend the analysis of the observability of gravitational decoherence due to time dilation to situations where memory effects play a crucial role. This is particularly relevant since reservoir engineering techniques nowadays allow to manipulate the properties of the environment in order to increase the coherence times, e.g., by means of backflow of information and recoherence characterizing non-Markovian dynamics \cite{RevOfNonMark}. This in turn would make it possible to prolong the coherence time associated to classical noise long enough to render time dilation induced decoherence practically observable in the experiments.

Our analysis follows the results of Ref. \cite{CollapseRevival}, where a quantum harmonic oscillator is subjected to a classical stochastic field as described by the following coupling Hamiltonian:

\begin{align}
H_{SC} =& \hbar \left( a \bar{B}(t) e^{i \omega t} + a^\dag B(t) e^{-i \omega t} \right),
\end{align}
where $B(t)=B_x(t)+i B_y(t)$ describes a Gaussian stochastic process with the following properties:
\begin{equation}
\begin{split}
\langle B_x(t) \rangle_B &= \langle B_y(t) \rangle_B = 0 \\
\langle B_x(t_1) B_x(t_2) \rangle_B &= \langle B_y(t_1) B_y(t_2) \rangle_B = K(t_1,t_2) \\
\langle B_x(t_1) B_y(t_2) \rangle_B &= \langle B_y(t_1) B_x(t_2) \rangle_B = 0.
\end{split}
\end{equation}
Here $\bar{B}$ is the complex conjugate of $B(t)$, the $\langle .. \rangle_B$ notation represents the average over all stochastic realizations, and $K(t_1,t_2)$ is the kernel autocorrelation function. For the sake of concreteness we consider a Ornstein-Uhlenbeck process with \cite{OrnsteinUhlebeck}
\begin{equation}
K(t_1,t_2) = \frac{1}{2} \lambda \gamma e^{- \gamma |t_1 - t_2|}.
\end{equation}

The parameter $\lambda$ is the system-noise coupling constant while $\gamma$ quantifies the temporal correlations of the environment, its inverse therefore measuring the so called memory time of the environment. For a Gaussian stationary process, the evolved state can be written in terms of the $s$-ordered characteristic functions as follows

\begin{equation}
\chi (\xi,s) = \chi_0 (\xi,s) e^{\frac{1}{2}|\xi|^2 \left( s - 2 \sigma(t) \right)}. \label{eq:classno}
\end{equation} 
with
\begin{equation}
\sigma(t) = \int_0^t \int_0^t ds_1 ds_2 \cos \left[ \delta (s_1-s_2) \right] K(s_1,s_2).
\end{equation}

%

For resonant interaction and for the Ornstein-Uhlenbeck process here considered,  $\sigma(t)$ has a simple analytical expression \cite{CollapseRevival}
\begin{equation}
\sigma(t) = \lambda t + \frac{\lambda}{\gamma} \left( e^{-\gamma t} - 1\right).
\end{equation}

Comparing Eq. (\ref{eq:classno}) with Eq. (\ref{eq:phi_with_ctdt}) one sees immediately that the quantum characteristic function describing the time evolution in presence of classical stochastic noise has precisely the same form as the one describing time dilation induced decoherence, with $\sigma (t)$ now playing the role of $N(t)$.

We recall here the expressions of the nonclassicality indicators calculated in \cite{CollapseRevival}. 
The decoherence time $\tau_W$, associated to the negativity of the Wigner function, is given by
\begin{equation}
t_W(\gamma,\lambda) = \omega_0 \left[ \frac{\gamma + 2 \lambda}{2 \gamma \lambda}+ \frac{1}{\gamma} \mathrm{ProductLog} \left( -e^{1-\frac{\gamma}{2 \lambda}} \right) \right] .
\end{equation}
As for the nonclassical depth it is sufficient to recall that, also in this case, $\tau_p^2 = 2 \tau_W^2$.
The Vogel and Klyshko criteria are studied according to the same lines of Sec. V. We note that, following Ref. \cite{CollapseRevival}, all energy-related quantities are rescaled in units of $\hbar \omega_0$.
\begin{figure}
\vspace{1cm}
\includegraphics[width=0.48\textwidth]{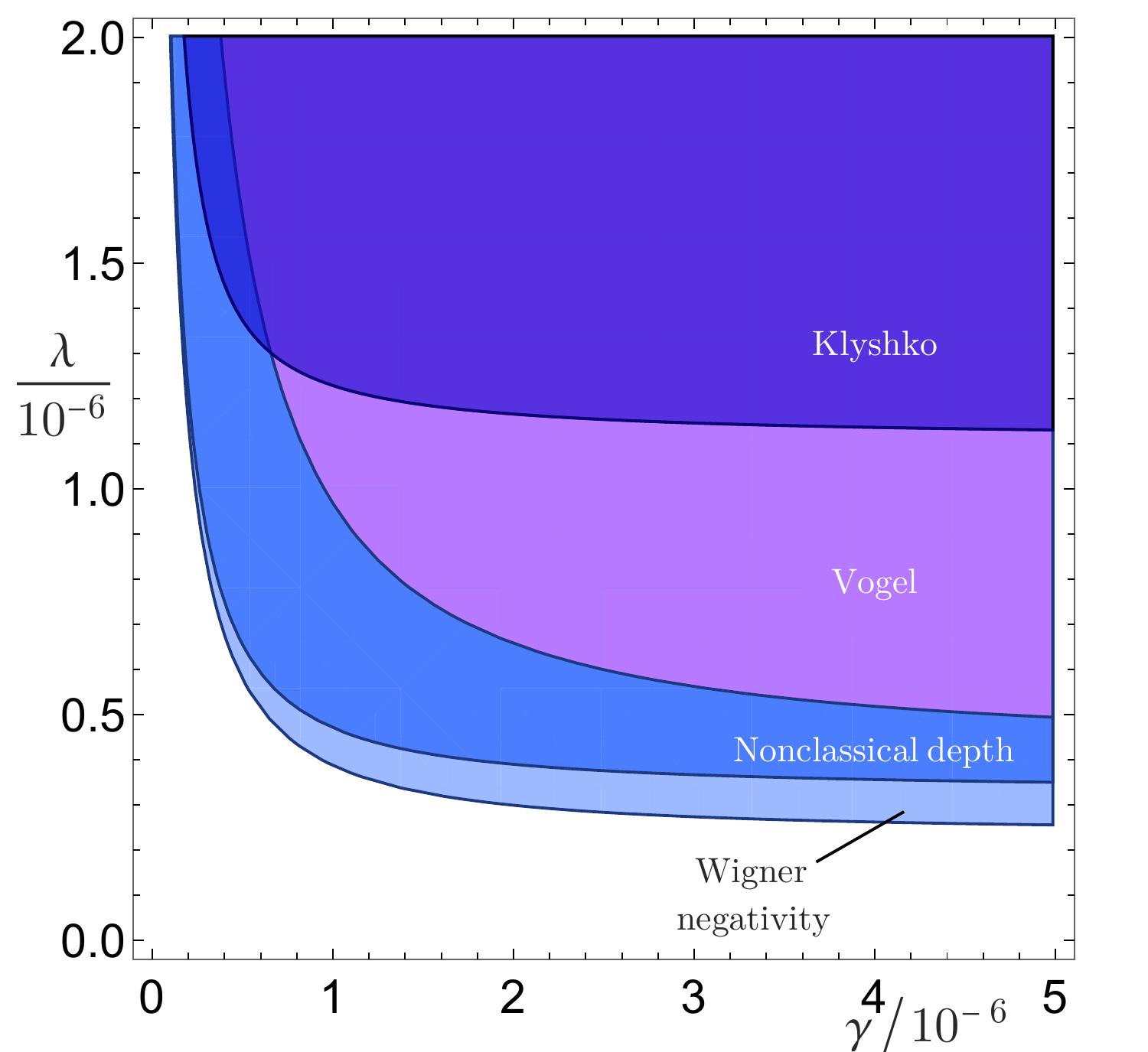}
\caption{(colors online) Ratio between the decoherence times for different nonclassicality indicators due to classical noise and those due to time dilation induced decoherence, as a function of $\gamma$ and $\lambda$ for $\alpha = \sqrt{2}$. The shaded areas represent the region of parameters in which  decoherence due to classical noise dominates over time dilation induced decoherence. $\gamma$ and $\lambda$ are given in units of $\omega_0$  (Hz).}\label{fig:classical_nondepth_wignerneg_vogel_klyshko}
\end{figure}
To compare the effects of classical noise and time dilation induced decoherence on the quantum to classical transition we consider the ratio of the respective decoherence times as a function of $\gamma$ and $\lambda$. Similarly to Ref. \cite{Brukner} we consider a system with superposition size of $\Delta x \approx 10^{-6}$ m and a temperature $T \approx 300 K$ setting the particle number $N$ to $\approx 10^5$. Regions in which the classical noise dominates are shown in Fig. \ref{fig:classical_nondepth_wignerneg_vogel_klyshko} and correspond to the ratio being $<1$. 
We notice that the behaviour of all different indicators with respect to the constants $\gamma$ and $\lambda$ is very similar and the constants' values must be on the order of $10^{-6}$ for the gravitational effect to be visible. This means that, independently of either the strength of the coupling or the memory time, and hence non-Markovian character of the dynamics, classical noise destroys all types of quantumness much faster than time dilation induced decoherence, making this phenomenon very difficult to observe in the experiments. 

Generally, estimates for the values of $\gamma$ and $\lambda$ are strongly dependent on the specific physical model and on the experimental implementation. Since no experiments have been performed with the physical system here considered, it is very difficult to predict the values of these two parameters. The most accurate experiment creating a Schr\"odinger cat state of motion of a material particle confined in a quantum harmonic trap was performed in the context of trapped ions \cite{DJWineland}. In that experiment the typical decoherence time, due to fluctuations of the trap potential, is of 10 $\mu$s. This corresponds to a value of $\gamma$ (in units of the trap frequency $\omega_0/2 \pi \approx 10^7$ Hz) of the order of $10^{-3}$, which is three orders of magnitude bigger than the bound we obtain. 


%

\section{Conclusions}

One of the goals of this paper was to investigate the potential ways to expand on the results of Ref. \cite{Brukner} concerning the decoherence of an open quantum system in a spatial superposition at different heights above a source of gravitational field. This was approached from several directions. Variable positions of system and observer manifest themselves in simple coefficients to the total decoherence time allowing for necessary corrections. We proved that the corrections are negligible close to earth surface but may become significant close to heavy stellar objects. We then considered the dynamics of a Schr\"odinger cat state and found the decoherence time measured in terms of the decay of the Wigner fringe visibility function. \\
In addition, we explored several other measures of decoherence, called nonclassicality indicators, and presented their dependence on the parameters of the system as well as their relative magnitudes. As some of the indicators are more appropriate for certain experimental implementations, the additional measures may prove useful for detection of the phenomenon. Finally, continuing with the theme of experimental detection, we examined the scale of precision or noise control required to detect the effect of time dilation in the presense of classical noise.  Our analysis shows that the phenomenon of gravitational induced decoherence imposes very high demands on the acceptable level of classical noise in order to be observed experimentally.

\section*{Acknowledgements} 

This work was supported by the Academy of Finland (Project no. 287750), Finnish  Cultural foundation and the Magnus Ehrnrooth Foundation.  We thank Prof. Daniel Braun for fruitful discussions.

\end{document}